# Learning an Integrated Distance Metric for Comparing Structure of Complex Networks

Sadegh Aliakbary, Sadegh Motallebi, Jafar Habibi, Ali Movaghar

Sharif University of Technology, Tehran, Iran

```
{aliakbary, motallebi}@ce.sharif.edu
     {jhabibi, movaghar}@sharif.edu
```

**Abstract.** Graph comparison plays a major role in many network applications. We often need a similarity metric for comparing networks according to their structural properties. Various network features – such as degree distribution and clustering coefficient – provide measurements for comparing networks from different points of view, but a global and integrated distance metric is still missing.

In this paper, we employ distance metric learning algorithms in order to construct an integrated distance metric for comparing structural properties of complex networks. According to natural witnesses of network similarities (such as network categories) the distance metric is learned by the means of a dataset of some labeled real networks. For evaluating our proposed method which is called NetDistance, we applied it as the distance metric in K-nearest-neighbors classification. Empirical results show that NetDistance outperforms previous methods, at least 20 percent, with respect to precision.

## 1 Introduction

Networks and graphs in real world appear in various forms. Social, biological, technological and information networks are just some examples of real networks. The graph representation of many real networks demonstrates nontrivial structural features that distinguish them from simple random graphs. Some of structural patterns in real networks are: sparseness, long-tail degree distributions, small worlds, transitivity of relationships and community structure. Such properties appear in many networks, but the degree of correspondence to these patterns is different for various network instances. For example, a network may display an obvious community structure while in another network the communities are not apparent.

In many applications, network comparison plays a major role and we frequently need a measure of similarity for comparing networks according to their structural properties. There are different measurements for calculating network similarities from the viewpoint of different structural properties. For example, we can compare two networks according to their density, clustering coefficient, degree distribution, average path lengths or any other structural measure. But many applications require a single integrated quantity as the overall similarity of two networks. Such an integrated measure has many applications in classification, clustering, model selection, anomaly detection and evaluation of sampling algorithms. As a simple approach for integrating different features in a single metric, we can create a feature vector for each network and then calculate the similarity of vectors according to methods such as Euclidean distance. We can also assign different weights for different properties manually, but we will show that it is better to apply machine learning methods based on witnesses of similarity/dissimilarity among real networks to achieve an integrated similarity metric.

In this paper, we propose a new network distance metric, which is named NetDistance. NetDistance compares the structure of the networks and returns the dissimilarity of network topological features. Our method has main differences with existing network distance metrics: we utilize machine learning for constructing the distance metric, we consider natural similarities of real networks and we integrate structural features in a single metric. We propose a methodology for learning an integrated distance metric in which "distance metric learning" algorithms are utilized for developing a distance metric for networks.

It is worth noting that when talking about network comparison, our intention is different from classical graph similarity approaches such as graph matching, graph isomorphism, graph alignment, edit distance and most of existing graph kernels. Such measurements do not reflect structural feature similarities of networks. They are also computationally expensive and inapplicable for large networks, such as those considered in this paper.

In the remainder of this paper, we interchangeably use phrases such as "real network", "complex network" or even "graph". Because "similarity" is the counterpart of "distance" or "dissimilarity" we may also use these terms for the meaning of quantified distance measurements for networks.

The structure of the rest of this paper is as follows: In section 2 we define the problem and its motivation. In section 3 we review the literature and related works. In section 4 the proposed method is illustrated. In the fifth section we evaluate our method and compare it with baseline methods. Finally, we conclude the paper in section 6.

## 2  Problem Definition and Motivation

The need for a structural similarity metric – i.e. the goal of this paper – is often discussed in the literature. Leskovec et al. [1] indicate the possibility of comparing the structure of different networks, even of different sizes, for calculating their similarity. They propose fitting a generative model (KronFit) for target networks and then using the differences in estimated parameters as a similarity measure. Sala et al. [2] also indicate "structure-driven graph comparison" and consider the problem of quantifying the similarity between any two graphs. The benefits of such a similarity metric is also discussed in [3][4].

In our problem setting, we aim a distance function that given two networks, calculates how similar they are. But what does this "similarity" mean? Some desired properties for the target similarity metric are following. For simplicity, we consider networks as simple graphs. A simple graph is an undirected, unweighted graph containing no self-loops or redundant edges.

- **Similarity in well-known structural features of complex networks**. Networks with similar values for structural measurements are considered similar. The integration of different features in a single metric is not a trivial task. The main question of this research is how to compute the distance of two networks when more than one of their features differ. In this context, other features (such as graphlets [3], aggregations of node features [5], etc.) may be useful, but our notion of similarity is based on well-known features of complex networks.
- **Independent from the size of the network.** The metric should have no assumption about the number of nodes and edges of the networks. An appropriate distance metric is able to compare networks of different sizes. If two networks have similar structural properties, whether they are similar in size or not, the metric should return a high level of similarity.
- **Similarity of networks of the same class**. We can categorize real networks in different classes such as friendship networks, citation networks, collaboration networks and etc. Networks of the same type are usually considered more similar than networks of different types. For example, two citation networks are probably more similar than a citation network and a friendship network. There is no standard class list for networks and even with a specified class list, the task of classifying a network may be nontrivial.
- **Generality**. The metric should be general enough to support different networks from different types. Even if we have not considered a specific class of networks in the learning process, the learned metric should be able to support the unseen classes.
- **Noise Tolerance**. Relatively small changes in networks should not result in large changes in network similarities. In this sense, methods such as edit distance are not considered noise tolerant.
- **Pseudometric for simple graphs**. We need the distance function to be a pseudometric d on N where N is the set of possible networks:
  - $d : N \times N \to \mathbb{R}$
  - $d(x, y) \geq 0$
  - $d(x, x) = 0$ (but possibly $d(x, y)=0$ for some distinct values x≠y)
  - $d(x, y) = d(y, x)$
  - $d(x, z) \leq d(x, y) + d(y, z)$

Our proposed method (NetDistance) supports the described requirements. Such a network distance metric has many applications in different domains such as evaluation of network generation models, classification and clustering, process prediction, evaluation of sampling methods, model selection, comparison of networks, an estimate of isomorphism and anomaly detection.

The art of NetDistance is to integrate complex network features in a single global metric. There are some measures for quantifying the structural features of networks. Here, we briefly review important topological network features that are incorporated in NetDistance:

- **Small-world**. In real networks most of the nodes can reach every other node by a small number of steps. Some measurements related to the path lengths are average shortest path length [6], radius [2], diameter [2] and effective diameter [7].
- **Heavy Tail Degree Distribution**. The degree distribution of many real networks follows a heavy-tailed and especially a power-law distribution. If we assume that the degree distribution of a network follows a power-law distribution, we can estimate the power-law exponent ($\gamma$) as a measurement of the degree distribution. But this assumption is rejected in some cases and the power-law degree distribution is not a good model for many social networks [8][9]. Statistical tests such as Kolmogorov-Smirnov test are also useful for comparing the degree distributions of two networks [10][11]. Calculating percentiles of the distribution is another approach for quantifying degree distributions as vectors of real numbers [3].
- **Sparseness**. Usually a small fraction of possible edges exist in real networks and the networks are considered sparse. Network density [8] and average degree [6] are measurements related to sparseness of networks.
- **Transitivity of Relationships**. Two nodes that are both neighbors of the same third node have more chance of also being neighbors of one another [12]. Clustering coefficient [13] and transitivity [6] are two well-known measures for quantifying the tendency of nodes for creating closed triads.
- **Community Structure**. The nodes of many real networks can be grouped in some clusters in such a way that the nodes in a cluster are more densely connected with each other than with the nodes of other clusters. Modularity [14] is one of the best measures for quantifying community structure of a network. Networks with high modularity have dense inter-community connections and sparse intra-community edges. In fact, the modularity of a network is dependent on the employed community detection algorithm.

There is no standard list for features and measurements of real networks and other patterns are also reported for real networks, such as degree correlation (assortativity) [15][16][17], densification [7], shrinking diameter [7], network resilience [17], vulnerability [18], navigability [19], and the rich-club phenomenon [20]. We have only considered simple graphs in this paper and measurements related to directed graphs (such as reciprocity [6]) are not investigated.

## 3 Literature Review

### 3.1 Distance Metric Learning

When data instances are represented by vectors of different features and there are also witnesses about similarity/dissimilarity of instances, we can learn a distance measure for the instances. "Distance Metric Learning" is the art of applying machine learning methods for finding a distance function for the input space of data from a given collection of similar/dissimilar instances and their corresponding features vectors.

Yang [21] has surveyed the field of distance metric learning along with its techniques and methods. Xing et al. [22] formulate the problem as a constrained convex programming problem. Weinberger et al. [23] show how to learn a Mahalanobis distance metric (called LMNN) for kNN classification from labeled examples. We found LMNN the most suitable algorithm for our application and we used it in this research.

### 3.2 Network Similarity Metrics

**Graph Isomorphism.** If two graphs have an identical topology we call them isomorphic. Because isomorphism is a strict metric, some variations for isomorphism are also presented in the literature such as subgraph isomorphism and maximum common subgraphs [24]. The degree of isomorphism between two graphs is measurable via metrics such as edit distance. There are also other methods for comparing graphs according to isomorphism criteria, such as counting their number of spanning trees [25], comparing graph spectrums [26] and computing similarity scores for nodes and edges [27]. Measures of this category are computationally expensive and they are totally inapplicable for large networks. In addition, they do not reflect structural properties of networks.

**Kernel Methods and Graph Kernels.** Kernel methods and graph kernels are also related to our research. A kernel $k(x, x')$ is a measure of similarity between objects $x$ and $x'$. Comparing graphs involves constructing a kernel for graphs and many kernels are proposed for graphs in the literature [28][29][30][31][24]. An appropriate kernel function should capture the characteristics of the graphs appropriately and it should also be efficiently computable and positive definite [29]. Graph kernels measure the similarity of graphs, but there is no common meaning for the term "similarity" in this domain. Sometimes kernel $K(G, H)$ is defined by counting subgraphs of G that have the same structure as graph H [30]. Gärtner et al. [30] defines the similarity of two graphs as a measure based on the length of all walks between each pair of nodes in the graph. Our research may be regarded as a graph kernel for comparing complex networks, but NetDistance differs from current graph kernel methods in some senses: we define "similarity" by the means of complex network features, we base our distance function on observations of real networks and we reach the distance function by the means of machine learning.

**Motif Counting.** Motif counting is an alternative approach for comparing networks. In order to characterize the local structure of graphs, it is possible to count some small subgraphs called motifs or graphlets. Motifs are small subgraphs and represent recurring, significant patterns of interconnections [32]. Similarity of motif counts may be used as a measure of similarity between graphs [3][32][33][34][35][36][37]. Motif counting is a computationally complex process and its methods are usually based on a pre-stage of network sampling [37][3].

**Feature Vectors.** Another family of distance measures for network comparison aims at representing the graphs by feature vectors that summarize graph topology [24]. The feature vector is usually called "topological descriptor" of the graph and its elements are well-known metrics of complex networks. In this approach the graph is replaced with a vector-representation and these vectors are utilized toward an indication of graph similarity. Many researches have mentioned feature vectors as a feasible method for graph comparison [3]. Some researchers have also utilized feature vectors for graph comparison and graph classification [5][38][4].

**Graph Signatures.** In some researches, a vector or a small matrix is extracted as the "signature" of the network and these signatures are used for network comparison, but the elements of this vector are not well-known network features [1][2]. Leskovec et al. [1] propose Kronecker graphs as a method for complex network generation. They propose KronFit as an algorithm for finding parameters (the initiator matrix K1) in Kronecker multiplication method for fitting an observed network. They list "graph similarity" as one of the applications of the initiator matrix. To compare the similarity of the structure of different networks (even of different sizes) one can use the differences in estimated parameters as a similarity measure. Sala et al. [2] propose to use the dK-series [39] as the similarity metric instead of focusing on known graph features. In dK-series method the d is usually limited to 2 because of inefficiency of the algorithm for larger values of d [2][39]. The running time of dK models and computation state size increase rapidly as d increases [2]. Airoldi et al. [40] propose a vector of 47 metrics as the feature vector. The vector contains mean, standard deviation, minimum and maximum of many node properties such as degree, clustering coefficient and etc.

## 4     Proposed Method

### 4.1    Methodology

Fig. 1 shows the methodology of this research. Instead of using naïve methods for network comparison (such as Euclidean distance) or developing manual or heuristic metrics, this methodology proposes to learn a network similarity metric based on natural similarities of networks in the real world. In fact, we have witnesses that some real networks are similar to each other and we can utilize this knowledge as training data for learning a similarity metric on networks.

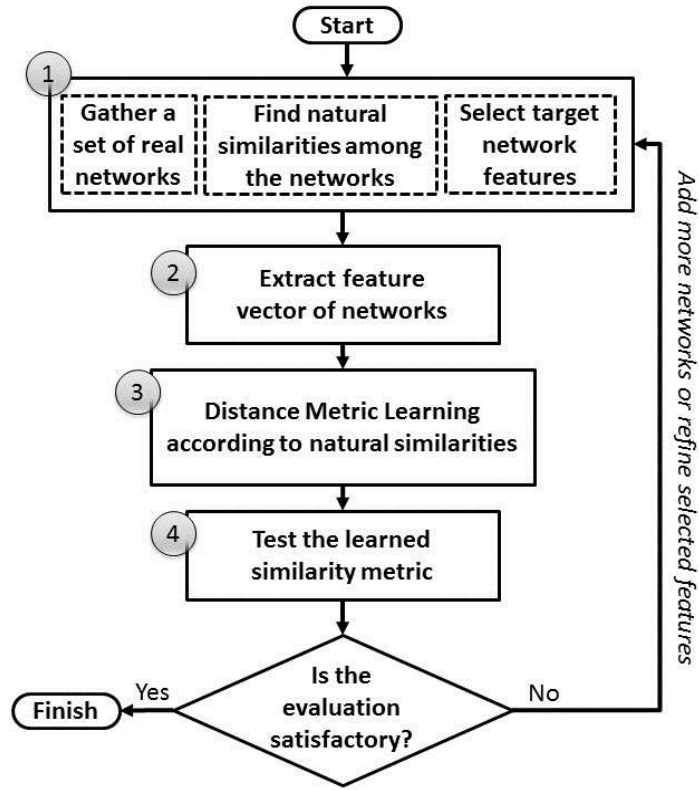

**Fig. 1.** - Methodology of learning a distance metric for complex networks

The proposed methodology includes the following steps:

1. Create and categorize a dataset and select the features.
    1.1. A set of real networks are gathered and represented as some simple graphs. Despite the existence of publicly available graphs of real networks, collecting a balanced and labeled set of networks is a difficult and costly task.
    1.2. Some witnesses about similarity/dissimilarity of selected networks are detected. We call the similarities that are discovered from the real world "natural similarities". For example, citation networks have natural similarity to each other. It is possible to realize natural similarities from network class labels, different versions of a network over time and different parts of a network. Since no standard exist for the set of network classes, careful selection of appropriate classes is needed so that enough real networks could be collected.
    1.3. Each network instance should be represented by a vector of features. The features are carefully selected among many complex network features that are illustrated earlier in this paper (degree distribution, average clustering, etc.)

2. Selected features are calculated for all the networks. Existing tools of network analysis may be used in this stage. The result is a feature vector for each network.
3. The set of feature vectors along with the natural similarities are used in a distance metric learning algorithm. The result is a metric for comparing structural properties of different networks. This metric specifies the weight of different network features for comparing networks.
4. Some instances of feature vectors should be detached, to be used in evaluation stage as the test-set. Obviously, the training-set and the test-set should be disjoint. Cross-validation should be employed to ensure the generality of learned metric. If the evaluation of the metric is not satisfactory, we should go back to the first stage and gather more networks or change the set of selected features or even the algorithm of distance metric learning.

### 4.2 Network Features

We explored important topological features of complex networks earlier in this paper. As our methodology specifies, we should select some of the features and form the feature vector of the networks. In this step, utilization of a wide and diverse feature set of network connectivity patterns is beneficial. As Table 1 shows, we have selected ten network features from the five main categories of topological features illustrated in the second section.

We have also designed a new method for quantifying the degree distribution of the networks. In this method, we extract some percentiles from the network distribution according to its mean and standard deviation. We devise $K$ intervals of equal size in the degree distribution and then we calculate the probability of degrees of each interval. The size of all intervals is considered equal to $p\sigma$ where $\sigma$ is the standard deviation of the distribution and $p$ is a coefficient that is determined by trial and error. The coefficient $p$ is tuned so that most of the node degrees lie in the created intervals. In our experiments we let $K = 4$ and $p = 0.3$, so we extract four quantities (DegDistP1..DegDistP4 percentiles) from any degree distribution. Formula 1 shows the interval points of degree distribution and Formula 2 specifies the probability for a node degree to sit in $i$th interval.

$$interval\_point_i = \mu - (\frac{K}{2} - i + 1)p\sigma \qquad (1)$$

$$DegDistP_i = P(degree > interval\_point_i \text{ AND } degree < interval\_point_{i+1}) \qquad (2)$$

The authors of [3] have pursued a similar approach for quantifying the degree distribution, but they break the distribution evenly into eight pieces and extract eight percentiles (called deg1..deg8). They do not consider the mean and the standard deviation of the distribution, so their specified intervals are biased toward outlier degree values. Our experiments showed that with the aim of comparing degree distributions, our method outperforms other methods such as percentiles proposed in [3], Kolmogorov-Smirnov test [10][11] and the power-law exponent.

**Table 1.** - Selected Topological Network Features

| Topological Feature | Selected Measurements |
|---|---|
| Small-world | Average Shortest Path |
| Degree Distribution | Four Percentiles (DegDistP1..DegDistP4) |
| Sparseness | Density, Average Degree |
| Transitivity of Relationships | Transitivity, Average Clustering |
| Community Structure | Modularity |

### 4.3 Dataset

We have collected a set of real networks from a wide range of network types. Most of the utilized networks are public and well-known datasets and many researches have utilized them. The networks are selected from six different complex network categories: friendship networks, communication networks, collaboration networks, citation networks, peer to peer networks and graph of linked web pages. The category of networks is a sign of natural similarity and the networks of the same type are considered more similar. Table 2 shows the set of real networks, which are used in the learning phase of this research.

### 4.4 Process of Learning

We propose to utilize distance metric learning algorithms for extracting the network distance function. We selected LMNN method [23] for learning the distance metric. LMNN method learns a pseudometric distance metric for kNN classification from labeled examples of the dataset. LMNN obtains a family of metrics by computing Euclidean distances after performing a linear transformation $\vec{x}' = L\vec{x}$. The distance metric is usually expressed in terms of the squared matrix $M$ which is defined in formula 3. If the elements of $L$ are real numbers, $M$ is guaranteed to be positive semidefinite. Formula 4 shows the squared distance in terms of the matrix $M$. In our experiments, LMNN outperformed former methods such as [22].

$$M = L^\mathsf{T} L \qquad (3)$$

$$D_M(\vec{x_i}, \vec{x_j}) = (\vec{x_i} - \vec{x_j})^\mathsf{T} M (\vec{x_i} - \vec{x_j}) \qquad (4)$$

We run LMNN algorithm with 5000 iterations and without dimensionality reduction option. We also normalize all features of the dataset before executing the learning algorithm. We use standard score (z-score) for normalization of features ($z = \frac{x-\mu}{\sigma}$). The dataset is always divided into disjoint sets of training and test data and $\mu$ and $\sigma$ are extracted from training data. In other words, test data have no effect on calculation of $\mu$ and $\sigma$ and we normalize the test set according to $\mu$ and $\sigma$ of the training data. To avoid over-fitting we perform multiple rounds of cross-validation with different partitions and the results are averaged over the rounds. We will show in the evaluation section that despite the relatively small size of the dataset, the learned metric is general, accurate, size-independent and independent of selected network categories.

**Table 2.** Dataset of real networks

| ID | Category | Vertices | Edges | Source |
|---|---|---|---|---|
| Cit-HepPh | Citation Network | 34,546 | 420,899 | SNAP[1] |
| Cit-HepTh | Citation Network | 27,770 | 352,304 | SNAP |
| dblp_cite | Citation Network | 475,886 | 2,284,694 | DBLP[2] |
| Cit_CiteSeerX | Citation Network | 1,106,431 | 11,791,228 | CiteSeerX[3] |
| CA-AstroPh | Collaboration Network | 18,772 | 198,080 | SNAP |
| CA-CondMat | Collaboration Network | 23,133 | 93,465 | SNAP |
| CA-HepTh | Collaboration Network | 9,877 | 25,985 | SNAP |
| CiteSeerX_Collaboration | Collaboration Network | 1,260,292 | 5,313,101 | CiteSeerX |
| com-dblp.ungraph | Collaboration Network | 317,080 | 1,049,866 | SNAP |
| dblp_collab | Collaboration Network | 975,044 | 3,489,572 | DBLP |
| refined_dblp20080824 | Collaboration Network | 511,163 | 1,871,070 | Sommer[4] |
| IMDB-USA-Commedy- | Collaboration Network | 4,155 | 16,679 | Rossetti[5] |
| CA-GrQc | Collaboration Network | 5,242 | 14,490 | SNAP |
| CA-HepPh | Collaboration Network | 12,008 | 118,505 | SNAP |
| Email | Communication Net- | 1,133 | 5,451 | Aarenas[6] |
| Email-Enron | Communication Net- | 36,692 | 183,831 | SNAP |
| Email-EuAll | Communication Net- | 265,214 | 365,025 | Konect[7] |
| WikiTalk | Communication Net- | 2,394,385 | 4,659,565 | SNAP |
| Dolphins | Friendship Network | 62 | 159 | NetData[8] |
| facebook-links | Friendship Network | 63,731 | 817,090 | MaxPlanck[9] |
| Slashdot0811 | Friendship Network | 77,360 | 507,833 | SNAP |
| Slashdot0902 | Friendship Network | 82,168 | 543,381 | SNAP |
| soc-Epinions1 | Friendship Network | 75,879 | 405,740 | SNAP |
| Twitter-Richmond-FF | Friendship Network | 2,566 | 8,593 | Rossetti |
| youtube-d-growth | Friendship Network | 1,138,499 | 2,990,443 | MaxPlanck |
| web-BerkStan | Graph of Web Pages | 685,230 | 6,649,470 | SNAP |
| web-Google | Graph of Web Pages | 875,713 | 4,322,051 | SNAP |
| web-NotreDame | Graph of Web Pages | 325,729 | 1,103,835 | SNAP |
| web-Stanford | Graph of Web Pages | 281,903 | 1,992,636 | SNAP |
| p2p-Gnutella04 | P2P Network | 10,876 | 39,994 | SNAP |
| p2p-Gnutella05 | P2P Network | 8,846 | 31,839 | SNAP |
| p2p-Gnutella06 | P2P Network | 8,717 | 31,525 | SNAP |
| p2p-Gnutella08 | P2P Network | 6,301 | 20,777 | SNAP |

---

[1] http://snap.stanford.edu/data
[2] http://dblp.uni-trier.de/xml/
[3] http://citeseerx.ist.psu.edu
[4] http://www.sommer.jp/graphs/
[5] http://www.giuliorossetti.net
[6] http://deim.urv.cat/~aarenas
[7] http://konect.uni-koblenz.de
[8] http://www-personal.umich.edu/~mejn/netdata/
[9] http://socialnetworks.mpi-sws.org

### 4.5 Implementation Issues

We have used several network tools and we have also implemented some necessary parts of this research. For calculating network features, we have used SNAP tool[10] (written by Jure Leskovec) and igraph library[11] of the R project. We used the public implementation of a new and fast algorithm proposed in [41] for community detection and calculating the modularity measure. The KronFit algorithm is also implemented and made available by Jure Leskovec. LMNN algorithm has a public MATLAB implementation[12]. We also implemented NetSimile and Euclidean distance methods in MATLAB. We have also utilized implementations of decision tree learning and support vector machines from Weka tool.

## 5 Evaluation

### 5.1 Baseline Methods

We compare the results with three baseline methods: NetSimile [5], KronFit [1] and Euclidean distance metric. Other methods are not involved in the evaluation because they are size-dependent, computationally inapplicable or out of scope in regard to our problem setting. We also compare the precision of our distance metric learning algorithm with the precision of decision tree learning and support vector machines.

NetSimile [5] proposes to extract a vector of 35 features from the network and then to compare networks according to Canberra distance of these vectors. NetSimile outperforms FSM (frequent subgraph mining) and EIG (eigenvalues extraction) methods [5]. We implemented NetSimile method and calculated the 35 features for all the networks of our dataset.

KronFit is the algorithm for fitting the Kronecker graph generation model to large real networks [1]. Leskovec et al. show that with KronFit, we can find a 2 × 2 initiator matrix (K1) that very well mimics the properties of the target network and using this matrix of four parameters we can accurately model several aspects of global network structure. They propose to compare the structure of networks (even of different sizes) by the means of the differences in estimated parameters. We calculated $K_1$ initiator matrix for all the networks of our dataset using KronFit method and we used its four features for comparing networks. Leskovec et al. do not specify the method of comparing initiator matrixes. So, among Canberra distance, Euclidean distance and average distance of feature vectors, we chose the best possible one in our evaluation and comparison.

NetSimile and KronFit methods try to form appropriate feature vectors for comparing networks. They do not utilize machine learning and indeed they assign an equal weight to all the features. We show that our method not only selects a better set of network features but also improves the distance metric by the means of machine learning. To demonstrate this claim, we involve the Euclidean distance of our selected features as another baseline method for comparing network structures. The evaluations show that this method outperforms KronFit and NetSimile. This fact shows that our proposed feature set is more appropriate than selected features of KronFit and NetSimile for summarizing network structure. In addition, our learning-based method (NetFistance) outperforms Euclidean distance method and it means that machine learning is able to improve the precision of the distance metric.

---
[10] http://snap.stanford.edu/snap/
[11] http://igraph.sourceforge.net/
[12] http://www.cse.wustl.edu/~kilian/

### 5.2 Comparison

NetDistance uses well-known network features (such as average path lengths and degree distributions) in the distance metric and this approach brings some benefits over other methods. In fact, the selected features convey special meanings and they have known consequences on graph behavior. The selected features are standard and popular network features and it is possible to calculate them with numerous existing software tools. They are already used as metrics of similarity in many researches, but we proposed to integrate these features in a single metric.

For evaluating the precision of NetDistance, we employ it, along with baseline methods, in kNN algorithm and we assess the precision of the resulting kNN classifier. kNN evaluation is a common approach for testing distance metric methods when we have a labeled dataset and we know the category of each record. Fig. 2 shows the precision of kNN classifier with different similarity metrics for various values of K. This figure shows that NetDistance has a steady and better precision in all situations. NetDistance is more accurate than Euclidean distance, so the learning stage is helpful in improving the precision of distance metric. Euclidean distance is better than KronFit and NetSimile and this fact shows that we have selected a proper set of features for comparing networks.

We averaged the precision of kNN for k=1, 3 and 4 and Table 3 shows the average precision of kNN classifier in different situations. The seventh and eighth rows of this table show the precision of other classifiers learned by the same dataset. In this phase of the evaluation, we have used implementations of J48 decision tree learning and SMO support vector machines. In the sixth row we show the precision of the algorithm without degree distribution features and it shows that degree distribution is an important feature and our method of quantifying degree distributions has improved the precision of the distance metric algorithm.

Someone may argue that our proposed distance metric is dependent on the set of chosen network categories and it is not indeed a general metric. To overcome this challenge, we evaluated the proposed method in a new manner. We limited the size of test-set to one instance and in each iteration of cross-validation, we eliminated all the classmates of the test-case (networks of the same category) from the training set. So, the classmates of the testcase are not involved in learning the distance metric but they are included in the set of instances in which kNN is applied. This setting shows whether the distance metric is able to correctly classify a test-case, even without learning from the same class of networks. The fifth row of Table 3 shows the average precision of kNN in this situation. This precision is still a high value (higher than

Euclidean distance method) and it shows the independence of the proposed distance metric from the set of selected network categories.

Table 4 also shows the average precision of kNN (with k=1, 3 and 4) separated by different network types. It shows that NetDistance not only outperforms other methods in average precision, but it is also better in detecting the class of the networks for most of the categories.

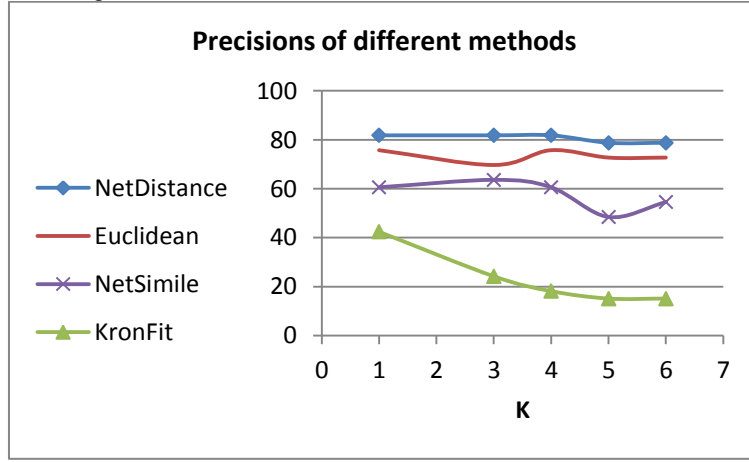

**Fig. 2.** - kNN evaluation of different network distance metrics with k=3,4 and 5.

**Table 3.** Precision of different distance metrics

|   | Distance Metric | Applied Change | Average Precision |
|---|---|---|---|
| 1 | Euclidean Distance | - | 73.74 |
| 2 | NetSimile | - | 61.62 |
| 3 | KronFit | - | 28.28 |
| 4 | *NetDistance* | - | *81.82* |
| 5 | NetDistance | Remove test-case classmates from training-set | 77.78 |
| 6 | NetDistance | Exclude features related to degree distribution | 76.77 |
| 7 | Decision Tree Learning (J48) | - | 58.15 |
| 8 | Support Vector Machines (SMO) | - | 69.62 |

**Table 4.** The precision of kNN separated by different network categories

| Category | #instances | NetDistance | Euclidean | KronFit | NetSimile | J48 | SMO |
|---|---|---|---|---|---|---|---|
| Friendship | 7 | 71.43 | 71.43 | 28.57 | 42.86 | 75 | 44.4 |
| Collaboration | 10 | 90.00 | 90.00 | 40.00 | 90.00 | 66.7 | 100 |
| Communication | 4 | 50.00 | 50.00 | 0.00 | 25.00 | 25 | 0 |
| Citation | 4 | 75.00 | 50.00 | 50.00 | 25.00 | 50 | 66.7 |
| Web Graph | 4 | 100.00 | 75.00 | 50.00 | 50.00 | 40 | 80 |
| P2P | 4 | 100.00 | 100.00 | 100.00 | 100.00 | 66.7 | 100 |
| Total | 33 | 81.82 | 75.76 | 42.42 | 60.61 | 58.15 | 69.62 |

The proposed method is also computationally more efficient than similar methods. In fact, the bottleneck of the distance metric algorithms is usually the calculation of the feature vectors and the comparison of feature vectors is usually a simple and fast stage. The feature vectors in NetDistance method consists of well-known network which are efficiently computable by the many network analysis tools. In our experiments, the feature vectors of NetDistance were computed so faster than the features of NetSimile and KronFit, especially for large networks. We have selected a diverse range of network sizes in our dataset. As a result, the final distance metric is independent of the size of the networks.

## 6      Conclusion and Future Works

This paper shows the possibility of learning a distance metric for comparing structures of complex networks. Our proposed distance metric (NetDistance) is a general and integrated metric that compares the network structures efficiently and precisely. We have used natural similarities from real world networks to learn the network distance metric.

The outcome of this paper, more than to be a static integrated measure, is a proof of concept for our proposed methodology. The selected features in feature vector of NetDistance convey special meanings and this is an opportunity for network simulation designers, because they can change the selected list of features according to the desired application. It is also possible to alter the dataset of networks and the classes of networks to improve the distance metric.

In future steps, we want to expand the records of the dataset and to create a larger dataset. We want to evaluate the noise tolerance of the proposed distance metric, because an appropriate distance measure should be insensitive to small network changes. We will also implement some network simulation scenarios to check whether structurally similar networks result in similar outcomes of simulations.

### Acknowledgements

The authors wish to thank Hossein Rahmani, Mehdi Jalili, Mahdieh Soleymani and Masoud Asadpour for helpful discussions and comments. Some of datasets of this work are prepared by Javad Gharechamani, Mahmood Neshati and Hadi Hashemi and we appreciate their cooperation and also those who prepare public network datasets and tools.